\documentstyle[11pt]{article}
\makeatletter
\renewcommand\section{\@startsection{section}{1}{\z@}%
  {-3.25ex\@plus -1ex \@minus -.2ex}%
  {0.8ex \@plus .2ex}{\reset@font\large\bfseries}}
\@addtoreset{equation}{section}
\makeatother

\def\be{\begin{equation}\label}            \def\ba{\begin{eqnarray}}
\def\ee{\end{equation}}                    \def\ea{\end{eqnarray}}
\def\ar{\begin{array}}   \def\er{\end{array}}   \def\nn{\nonumber} 
\def\ts{\textstyle}             
    \def\={\,=}
\begin{document}
\setcounter{page}{0}
\thispagestyle{empty}
{\small\sc April~~1999 \hfill  hep-th/9904059 }
\vspace*{1.7cm}

\begin{center}
{\large\bf Fermionic representations for characters of \\
 ${\cal M}(3,t)$,  ${\cal M}(4,5)$, ${\cal M}(5,6)$ and ${\cal M}(6,7)$
 minimal models and \\ 
 related Rogers-Ramanujan type and dilogarithm identities} \\ [1.1cm]
{\sc Andrei G. Bytsko}
  \\ [4mm]
Steklov Mathematics Institute, Fontanka 27,\\
 St.Petersburg~~191011, Russia \\ [20mm]
\hfill \begin{tabular}{c}
 Dedicated to Professor L.D.Faddeev \\ on his 65th birthday. 
 \end{tabular} 
\\ [40mm]
{\bf Abstract} \\ [9mm]

\parbox{13cm}{ 
 Characters and linear combinations of characters that admit a fermionic
 sum representation as well as a factorized form are considered for some 
 minimal Virasoro models. 
 As a consequence, various Rogers-Ramanujan type identities are obtained.
 Dilogarithm identities producing corresponding effective central charges
 and secondary effective central charges are derived.
 Several ways of constructing more general fermionic representations are
 discussed.
}  \end{center}

\vspace*{2cm}


\vfill{ \hspace*{-9mm}
 \begin{tabular}{l}
\rule{6 cm}{0.05 mm} \\
 bytsko@pdmi.ras.ru 
 \end{tabular} }
\setcounter{section}{0}

\newpage

\section{Introduction} 

A minimal Virasoro model \cite{BPZ} ${\cal M}(s,t)$ is parameterized by 
two positive integers $s$ and $t$ such that $\langle s, t \rangle =1$ 
(i.e.~they are co-prime numbers). It has the central charge
$c(s,t) = 1 - \frac{6(s-t)^{2}}{st}$. 
The characters of its irreducible representations with highest weights 
$h_{n,m}^{s,t}=\frac{(nt-ms)^2-(s-t)^2}{4st}$ are given by \cite{FF}
\be{char}
 \chi_{n,m}^{s,t}(q) \= \frac{ q^{\eta_{n,m}^{s,t}} }{(q)_\infty }
  \sum_{k=-\infty }^{\infty }q^{st k^2}
 \left(q^{k(nt-ms)}-q^{k(nt+ms)+nm}\right) \, ,
\ee
where $1{\leq} n{\leq} s{-}1$, $1{\leq} m{\leq} t{-}1$,\, 
$\eta_{n,m}^{s,t}:={h_{n,m}^{s,t} - \frac{c(s,t)}{24}}$ and
$(q)_m:=\prod_{k=1}^m (1-q^k)$. 
The characters possess the following symmetries
\be{hsym} 
 \chi^{s,t}_{n,m}(q) \= \chi^{t,s}_{m,n}(q) \= 
 \chi^{s,t}_{s-n,t-m}(q)  \= \chi^{t,s}_{t-m,s-n}(q) \, .
\ee
In addition, (\ref{char}) allows us to relate some characters of 
different models
\be{alp}
   \chi^{\alpha s,t}_{\alpha n,m}(q) \= 
   \chi^{s, \alpha t}_{n, \alpha m}(q) \, ,
\ee
where $\alpha$ is a positive number such that
$\langle s, \alpha t \rangle = \langle t, \alpha s \rangle =1$. 
For instance, $ \chi^{5,6}_{m,2}(q) = \chi^{3,10}_{1,2m}(q)$, $m=1,2$. 
Below we will also use the identity proven in \cite{BF2}:
\be{36}
 \chi^{3n,2t}_{n,t-2m}(q) - \chi^{3n,2t}_{n,t+2m}(q) \=
 \chi^{6n,t}_{n,m}(q) - \chi^{6n,t}_{n,t-m}(q) \,,
\ee
where $m< t/2$ and
$\langle t,6 \rangle = \langle n,2\rangle = \langle t,n\rangle =1$.

In some cases characters (\ref{char}) admit the form named ``fermionic
representation"
\be{F1}
 \chi(q) \= q^{\rm{const}} \sum\limits_{\vec{m}=\vec{0}}^\infty  
 \frac{q^{\vec{m}^{\,t} A \vec{m} + \vec{m}\cdot\vec{B} }}
 {(q)_{m_{1}}\ldots (q)_{m_{r}}}\,.
\ee
Hitherto examples of such representation were obtained for two large
classes of characters. For $A$ being related to the inverse Cartan matrix 
for some Lie algebra they were studied in \cite{KKMM}. In this case
certain restrictions are often imposed on the summation over $\vec{m}$. 
Examples of (\ref{F1}) for $A=0$ or, at most, being a diagonal matrix 
(and $(q)_{m_i}$ being replaced with 
$(q^b)_{m_i}$, $b>0$) were obtained in \cite{BF2,BF1} as 
consequences of representation of characters in the ``factorized form"
\be{fact}
 \chi(q) \= q^{\rm{const}}  
 \prod_{m=1}^M \left(\{x_m\}^+_y\right)^{\gamma^+_m}
 \prod_{n=1}^N \left(\{x_n\}^-_y\right)^{\gamma^-_n} \,,
\ee
where the multiplicities $\gamma^\pm_i$ are integer. Here and below
we use the notations of \cite{BF2} 
\be{bl}
\{x\}_{y}^{\pm }:=\prod\limits_{k=0}^{\infty }\left( 1\pm q^{x+ky}\right)
\,,\quad0<x<y\,;\quad\quad 
\left\{ x_{1};\ldots ;x_{n}\right\}_{y}^{\pm}:=
\prod\limits_{a=1}^{n}\{x_{a}\}_{y}^{\pm }\,.
\ee
Equivalence of fermionic and factorized forms for characters which
admit both types of representation gives rise to non-trivial
identities. We will refer to them as to Rogers-Ramanujan type identities.

The paper is organized as follows. Section 2 contains fermionic sum
representations for certain families of characters and linear combinations
of characters for ${\cal M}(3,t)$. These results are extensions of some 
previously known examples. Here are also given several fermionic sum 
representations for ${\cal M}(4,5)$, ${\cal M}(5,6)$ and ${\cal M}(6,7)$ 
that seem to be new. In Section 3 we observe that all the considered 
(combinations of) characters admit also the factorized form (\ref{fact})
and present the corresponding Rogers-Ramanujan type identities. 
In Section 4 we apply the saddle point analysis to the fermionic 
representations given in Section 2 and derive Bethe ansatz type equations 
that yield the corresponding effective central charge or (for the 
differences of characters) the secondary effective central charge as a 
sum of dilogarithms. For the latter case we show how to modify the saddle 
point analysis if we are dealing with fermionic sum with alternated signs
in summation. We solve the Bethe ansatz type equations explicitly and 
obtain four infinite dilogarithm sum rules corresponding to ${\cal M}(3,t)$
and one non-trivial identity for ${\cal M}(4,5)$. In Section 5 we employ 
certain identities for Virasoro characters and expand the list of fermionic 
representations for ${\cal M}(4,5)$, ${\cal M}(5,6)$ and ${\cal M}(6,7)$.
Section 6 contains a brief discussion and conclusion.

\section{Fermionic representations for ${\cal M}(3,t)$, ${\cal M}(4,5)$
  ${\cal M}(5,6)$ and ${\cal M}(6,7)$.} 

It was observed in \cite{KKMM} that the characters 
$\chi^{3,3k+1}_{1,k}(q)$ and $\chi^{3,3k+2}_{1,k}(q)$ as well as
all the four characters of ${\cal M}(3,5)$ admit the fermionic form
(\ref{F1}) with $A_k$ and ${\tilde A}_k$ given by
\ba
&& \bigl(A_k\bigr)_{ij} = \bigl(A_k\bigr)_{ji} = \min (i,j) \,, 
 \quad {\rm for}\ 1\leq i,j \leq k-1 \,,  \label{A3a} \\
&& \bigl(A_k\bigr)_{kj} = \bigl(A_k\bigr)_{jk} = 
 {\ts \frac j2} + {\ts \frac{1-k}{4}} \delta_{jk} \,,
 \label{A3b} \\
&& \bigl({\tilde A}_k\bigr)_{ij} \= \bigl(A_k\bigr)_{ij} -
{\ts \frac 14 } \delta_{ik} \delta_{jk} \,. \label{A3c}
\ea
It turns out that these results can be extended as follows:
\ba
 \chi^{3,3k+1}_{1,n}(q) \pm \chi^{3,3k+1}_{1,3k+1-n}(q) &=&
 q^{\eta^{3,3k+1}_{1,n}} \sum\limits_{\vec{m}=\vec{0}}^\infty 
 \frac{(\pm 1)^{m_k} q^{\vec{m}^{\,t} A_k \vec{m} + 
 \vec{m}\cdot\vec{B}^{3k+1}_{n}}}{ 
 (q)_{m_{1}} \ldots (q)_{m_{k}}} \, , \label{M31} \\
 \chi^{3,3k+2}_{1,n}(q) \pm \chi^{3,3k+2}_{1,3k+2-n}(q) &=&
 q^{\eta^{3,3k+2}_{1,n}} \sum\limits_{\vec{m}=\vec{0}}^\infty 
 \frac{(\pm 1)^{m_k}  q^{\vec{m}^{\,t} {\tilde A}_k \vec{m} + 
 \vec{m}\cdot\vec{B}^{3k+2}_{n} }}
 {(q)_{m_{1}} \ldots (q)_{m_{k}}} \, , \label{M32}
\ea
where $1{\leq} n {\leq} (k{+}1)$, the matrices $A_k$ and ${\tilde A}_k$ 
are defined by (\ref{A3a})-(\ref{A3c}) and the corresponding $k$-component
vectors ${\vec B}^{3k+1}_{n}$ and ${\vec B}^{3k+2}_{n}$ are such that
\ba
 && \label{B3} \bigl( {\vec B}^{3k+1}_{n} \bigr)_j \=
 \max (j-n+1,0) + {\ts \frac{n-k-2}{2}} \delta_{jk} \,,  \\
 && \label{B4} \bigl( {\vec B}^{3k+2}_{n} \bigr)_j \= 
 \bigl({\vec B}^{3k+1}_{n} \bigr)_j +{\ts \frac 12}\delta_{jk} \,.
\ea
For example, 
${\vec B}^{3k+1}_{1}=(1,2,3,\ldots,k-1,{\ts \frac{k-1}{2}})$,
${\vec B}^{3k+1}_{k}={\vec 0}$ and ${\vec B}^{3k+1}_{k+1} =
(0,\ldots,0,{\ts -\frac 12})$.

It turns out that besides the infinite families (\ref{M31}) and (\ref{M32})
of fermionic sums for the combinations of characters there exist similar
ones for ``single" characters:
\ba   
 \label{M33}  && \chi^{3,6t-2}_{1,3t-1}(q) = 
 q^{\eta^{3,6t-2}_{1,3t-1}} \sum\limits_{\vec{m}=\vec{0}}^\infty 
 \frac{ q^{\vec{m}^{\,t} A_{2t-1} \vec{m} + 
 \vec{m}\cdot\vec{C}^{2t-1}}}{(q)_{m_{1}} \ldots (q)_{m_{2t-1}}} \,,\\
 \label{M34}  && \chi^{3,6t+2}_{1,3t+1}(q) = 
 q^{\eta^{3,6t+2}_{1,3t+1}} \sum\limits_{\vec{m}=\vec{0}}^\infty 
 \frac{ q^{\vec{m}^{\,t} {\tilde A}_{2t} \vec{m} + 
 \vec{m}\cdot\vec{C}^{2t} }} {(q)_{m_{1}} \ldots (q)_{m_{2t}}} \, , 
\ea
where $t=1,2,3\ldots$, $A$ and ${\tilde A}$ are defined in 
(\ref{A3a})-(\ref{A3c}), and
\be{M35}
 \bigl(\vec{C}^{2t-1}\bigr)_j  \=  \max (j-t+1,0) - 
  {\ts \frac t2} \delta_{j,2t-1} \,,\quad \quad
\bigl(\vec{C}^{2t}\bigr)_j  \=  \max (j-t+1,0) - 
  {\ts \frac{t+2}{2} } \delta_{j,2t} \,. 
\ee
For example, $\vec{C}^1=(\frac 12)$, $\vec{C}^2=(1,\frac 12)$,
$\vec{C}^2=(0,1,1)$, $\vec{C}^3=(0,1,2,1)$.

Although formal derivation of (\ref{M31})-(\ref{M32}) and 
(\ref{M33})-(\ref{M34}) is beyond 
the scope of the present paper, I have verified these identities using
Mathematica for $k\leq 7$, $t\leq 5$ expanding them typically up to 
$q^{100}$. A rigorous proof can presumably be achieved with the help 
of the machinery of Bailey pairs (see e.g.~\cite{BP}). In contrast to 
a generic fermionic representation, eqs.~(\ref{M31}) and (\ref{M32})
do not have restrictions on the summation. However, if we combine
them to obtain an expression for a single character, the summation 
over $m_k$ will be restricted to either odd or even numbers. In particular,
eq.~(\ref{M31}) in the case of $k=1$, $n=2$ gives two expressions
for $\chi^{3,4}_{1,2}(q)$ (as noted also in \cite{KKMM}). Moreover,
eq.~(\ref{M33}) yields one more expression for the same character 
and hence we get 
\be{34}
 q^{-\eta^{3,4}_{1,2}} \chi^{3,4}_{1,2}(q) \= \sum_{m\geq 1 
 \atop m-{\rm odd}} \frac{ q^{\frac 12 m^2-\frac 12 m} } {(q)_m} = 
 \sum_{m\geq 0 \atop m-{\rm even}} \frac{ q^{\frac 12 m^2-\frac 12 m} }
 {(q)_m} =\sum_{m=0}^\infty \frac{q^{\frac 12 m^2+\frac 12 m}}{(q)_m} \,.
\ee 

While (\ref{M31}) and (\ref{M32}) extend previously known examples,
the following fermionic representations for ${\cal M}(4,5)$, 
${\cal M}(5,6)$ and ${\cal M}(6,7)$ to my knowledge have not been 
discussed in the literature so far:

\noindent
\underline{${\cal M}(4,5)$:}\  
two characters are representable in the form (\ref{F1}) with 
\be{M45a}
  A \= {\ts \frac 12}
 \left( \ar{cc} 4 & 1 \\ 1 & 1 \er\right)\, ,
\ee
namely, for $n=1,2$ the following equality holds
\be{M45b}
 \chi^{4,5}_{2,n}(q) \= q^{\eta^{4,5}_{1,n}} \sum_{\vec{m}=\vec{0}}^\infty 
 \frac{ q^{2m_1^2+\frac 12 m_2^2+m_1m_2+(4-2n)m_1+ \frac 12 m_2} }
 { (q)_{m_1} (q)_{m_2} } \,.
\ee

\noindent 
\underline{${\cal M}(5,6)$:}\ certain linear combinations of characters
admit the form (\ref{F1}) with 
\be{M56a}
  A \={\ts \frac 12} \left( \ar{cc} 1 & 1 \\ 1 & 1 \er\right)\,,
\ee
namely, for $n=1,2$, we have
\ba
 && \chi^{5,6}_{n,2}(q) \pm \chi^{5,6}_{n,4}(q) \=q^{\eta^{5,6}_{n,2}}
 \sum_{\vec{m}=\vec{0}}^\infty \frac{ (\pm 1)^{m_2} q^{\frac 12 (m_1^2
 + m_2^2) +m_1m_2+ \frac 12 m_1+(2- n)m_2} }
 { (q)_{m_1} (q)_{m_2} } \,, \label{M56b} \\
 && \chi^{5,6}_{n,1}(q) - \chi^{5,6}_{n,5}(q) \=q^{\eta^{5,6}_{n,1}}
 \sum_{\vec{m}=\vec{0}}^\infty  \frac{ (-1)^{m_2} q^{\frac 12 (m_1^2 +
 m_2^2) +m_1m_2+ (n-1)(m_1+m_2) } }
 { (q)_{m_1} (q)_{m_2} } \,. \label{M56c} 
\ea
\underline{${\cal M}(5,6)$:}\ Another fermionic representation for
the characters of the ${\cal M}(5,6)$ model can be obtained if we
notice that eqs.~(\ref{alp}) and (\ref{36}) allow us to relate
these characters to those of the ${\cal M}(3,10)$ model:
$ \chi^{5,6}_{n,2}(q) \pm \chi^{5,6}_{n,4}(q)=
\chi^{3,10}_{1,2n}(q) \pm \chi^{3,10}_{1,10-2n}(q)$ and  
$ \chi^{5,6}_{n,1}(q) - \chi^{5,6}_{n,5}(q)=
\chi^{3,10}_{1,5-2n}(q) - \chi^{3,10}_{1,5+2n}(q)$, $n=1,2$. 
Combining these relations with formulae (\ref{M31}) 
for $k=3$, we find
\ba
 && \chi^{5,6}_{n,2}(q) \pm \chi^{5,6}_{n,4}(q) \=q^{\eta^{5,6}_{n,2}}
 \sum_{\vec{m}=\vec{0}}^\infty\frac{ (\pm 1)^{m_3} q^{\vec{m}^t A \vec{m}
 + (2-n)m_2 + (\frac 32 -n)m_3} }
 { (q)_{m_1} (q)_{m_2} (q)_{m_3}} \,, \label{M56d} \\
 && \chi^{5,6}_{n,1}(q) - \chi^{5,6}_{n,5}(q) \=q^{\eta^{5,6}_{n,1}}
 \sum_{\vec{m}=\vec{0}}^\infty \frac{ (-1)^{m_3} q^{ \vec{m}^t A \vec{m}+
 (n-1)(m_1+2m_2 +m_3)} }
 { (q)_{m_1} (q)_{m_2} (q)_{m_3}} \,, \label{M56e} 
\ea
where $n=1,2$ and
\be{M56f}
  A \= \left( \ar{ccc} 1 & 1 & {\ts \frac 12} \\
  1 & 2 & 1 \\ {\ts \frac 12} & 1 & 1 \er\right)\, .
\ee
Furthermore, eq.~(\ref{alp}) also implies that 
$ \chi^{5,6}_{n,3}(q) = \chi^{2,15}_{1,3n}(q)$, $n=1,2$. For the
${\cal M}(2,2k+1)$ model a fermionic representation is well known
\cite{And,Kac,NRT}:
\be{2t}
 \chi^{2,2k+1}_{1,n}(q) \= q^{\eta^{2,2k+1}_{1,n}} \sum\limits_{\vec{m}
 =\vec{0}}^\infty \frac{q^{\vec{m}^{\,t} {\bar A}_k \vec{m} + 
 \vec{m} \cdot \vec{F}^k_n }}{(q)_{m_{1}}\ldots (q)_{m_{k-1}}}\,,
\ee
where $1{\leq}n{\leq}k$,\, $\bigl(\vec{F}^k_n\bigr)_j=\max(j-n+1,0)$,
and ${\bar A}_k$ coincides with the $\textstyle (k{-}1){\times}(k{-}1)$
minor (\ref{A3a}) which is the inverse Cartan matrix of the tadpole graph 
with $(k{-}1)$ nodes. This gives us yet another fermionic representation 
for ${\cal M}(5,6)$ with $A$ being 6$\times$6 matrix.

\noindent
\underline{${\cal M}(6,7)$:}\ Employing again eqs.~(\ref{alp}) and 
(\ref{36}), we observe that
$ \chi^{6,7}_{2,n}(q) \pm \chi^{6,7}_{4,n}(q)=
\chi^{3,14}_{1,2n}(q) \pm \chi^{3,14}_{1,14-2n}(q)$ and
$ \chi^{6,7}_{1,n}(q) - \chi^{6,7}_{5,n}(q)=
\chi^{3,14}_{1,7-2n}(q) - \chi^{3,14}_{1,7+2n}(q)$,
where $n=1,2,3$. Combining these relations with formulae (\ref{M32}) 
for $k=4$, we find
\ba
 && \!\!\!\!\!\!\!\!\!\!\! \chi^{6,7}_{2,n}(q) \pm \chi^{6,7}_{4,n}(q) = 
 q^{\eta^{6,7}_{2,n}}  \sum_{\vec{m}=\vec{0}}^\infty \frac{ (\pm 1)^{m_4}
 q^{\vec{m}^t A \vec{m} + (2-n)(m_2 +2m_3) + (\frac 52-n)m_4} }{(q)_{m_1}
 (q)_{m_2} (q)_{m_3} (q)_{m_4}} ,\quad n=1,2 , \label{M67a} \\
 && \!\!\!\!\!\!\!\!\!\!\! \chi^{6,7}_{1,n}(q) - \chi^{6,7}_{5,n}(q) = 
 q^{\eta^{6,7}_{1,n}} \sum_{\vec{m}=\vec{0}}^\infty \frac{ (-1)^{m_4}
 q^{\vec{m}^t A \vec{m} + \vec{m}\cdot\vec{D}^n} }{ (q)_{m_1} (q)_{m_2} 
 (q)_{m_3} (q)_{m_4} } ,\quad n=1,2,3 ,\label{M67c}
\ea
where $\vec{D}^1=(0,0,0,0)$, $\vec{D}^2=(0,0,1,1)$, $\vec{D}^3=(1,2,3,2)$
and 
\be{M67d}
  A \= \left( \ar{cccc} 1& 1 & 1 & {\ts \frac 12} \\
  1 & 2 & 2 & 1 \\ 1& 2& 3 &{\ts \frac 32}\\ 
  {\ts \frac 12} & 1 & {\ts \frac 32}& 1 \er\right)\, .
\ee
Furthermore, due to eq.~(\ref{alp}) we can identify
$ \chi^{6,7}_{3,n}(q) = \chi^{2,21}_{1,3n}(q)$, $n=1,2,3$. Therefore,
these three characters of ${\cal M}(6,7)$ admit fermionic form
of the type (\ref{2t}) with $A$ being 9$\times$9 matrix.

\section{Rogers-Ramanujan type identities.} 

In the previous section we considered some characters and combinations
of characters which possess fermionic representations. It turns out that
all of them have another common feature -- they are factorizable, that 
is they admit also the form (\ref{fact}). For the characters of the
${\cal M}(2,2k+1)$ models this is the well-known representation
\be{AG}
 \chi^{2,2k+1}_{1,n}(q) \=  q^{\eta^{2,2k+1}_{1,n}} 
 \prod_{{j=1} \atop{j\neq n}}^{k} \frac{1}{ \{j;2k+1-j\}^-_{2k+1} } \,,
\ee
where we use the notations (\ref{bl}). Equality of the r.h.s.~of
eqs.~(\ref{AG}) and (\ref{2t}) yields a family of identities known as 
the Andrews-Gordon identities \cite{And}. For $k=2$ these are the
famous Rogers-Ramanujan identities \cite{RSR}
\be{RR}
 \sum_{m=0}^\infty \frac{q^{m^2+m}}{(q)_m} \=
 \frac{1}{ \{2;3\}_{5}^{-} }  \,, \quad\quad
 \sum_{m=0}^\infty \frac{q^{m^2}}{(q)_m} \=
 \frac{1}{ \{1;4\}_{5}^{-} }   \,.
\ee

Actually, (\ref{AG}) is only a particular case of a more general
formula 
\be{2s}
 \chi_{n,m}^{2n,t}(q) \= \frac{q^{\eta^{2n,t}_{n,m} }}{\{1\}^-_1} 
 \{nm;nt-nm;nt\}_{nt}^- \, ,
\ee
which together with
\be{3s}
 \chi_{n,m}^{3n,t}(q) \= \frac{q^{ \eta^{3n,t}_{n,m} }}{\{1\}^-_1}
 \{nm;2nt-nm;2nt\}_{2nt}^- \{2nt-2nm;2nt+2nm\}_{4nt}^{-}  
\ee
exhausts the possibility for single characters to be factorizable
on the base of the $A_1^{(1)}$ and $A_2^{(2)}$ Macdonald identities
\cite{BF2,Chr,CIZ}. Furthermore, it was shown in \cite{BF2} that 
in certain cases the following combinations
\be{pm}
 \chi_{n,m}^{s,t}(q) \pm \chi_{n,t-m}^{s,t}(q)
\ee
also are factorizable on the base of the same Macdonald identities.
The explicit formulae found in \cite{BF2} read
\begin{eqnarray}
 \chi_{n,m}^{3n,t}(q) \pm \chi_{2n,m}^{3n,t}(q) &=&
 \frac{q^{\eta^{3n,t}_{n,m} }}{ \{1\}^-_1} \{ nm;nt-nm\} _{nt}^-
{\ts \left\{ \frac{nt}{2}\right\}_{\frac{nt}{2}}^- 
\left\{ \frac{nt-2nm}{4}; \frac{nt+2nm}{4}\right\}_{\frac{nt}{2}}^{\pm} }
 \,, \label{f3} \\
 \chi_{n,m}^{4n,t}(q) \pm \chi_{3n,m}^{4n,t}(q) &=&
 \frac{q^{\eta^{4n,t}_{n,m} }}{ \{1\}^-_1} \{nm;nt-nm;nt\} _{nt}^{-}  
 {\ts \left\{ \frac{nt}{2}-nm;\frac{nt}{2}+nm;
 \frac{nt}{2}\right\}_{nt}^\pm } \, ,  \label{f4} \\
 \chi_{n,m}^{6n,t}(q) - \chi_{5n,m}^{6n,t}(q) &=&
 \frac{q^{\eta^{6n,t}_{n,m} }}{ \{1\}^-_1} \, \{nm;nt-nm;nt\}_{nt}^-
 \{ nt-2nm;nt+2nm\}_{2nt}^-   \,.  \label{f6}
\end{eqnarray}
Combining the fermionic representations given in the previous section
with these factorized representations, we obtain various identities of 
the ``sum--product" type. They can be regarded as generalizations of 
the Rogers-Ramanujan identities and it seems that many of them  
(especially those with a multivariable summation) have not appeared in 
the literature so far. 

\noindent
\underline{${\cal M}(3,t)$:}\ According to (\ref{f3}) the fermionic sums
on the r.h.s.~of (\ref{M31})-(\ref{M32}) are equal, respectively, to
\ba
 \frac{q^{\eta^{3,3k+1}_{1,n}}}{ \{1\}^-_1} \{ n;3k+1-n\} _{3k+1}^-
 {\ts \left\{ \frac{3k+1}{2}\right\}_{\frac{3k+1}{2}}^- 
 \left\{\frac{3k+1-2n}{4}; \frac{3k+1+2n}{4}\right\}_{\frac{3k+1}{2}}^{\pm}
 } \,,\label{rr1}  \\
 \frac{q^{\eta^{3,3k+2}_{1,n}}}{ \{1\}^-_1} \{ n;3k+2-n\} _{3k+2}^-
 {\ts \left\{ \frac{3k+2}{2}\right\}_{\frac{3k+2}{2}}^- 
 \left\{\frac{3k+2-2n}{4}; \frac{3k+2+2n}{4}\right\}_{\frac{3k+2}{2}}^{\pm}
 } \,.\label{rr2} 
\ea
For instance, for $k=1,2$ we get the following identities
\ba
 && \!\!\!\!\!\!\!\!\!\!\!\!  
 \sum_{m=0}^\infty  \frac{ (\pm 1)^{m}\, q^{ \frac 12 m^2}}
 {(q)_m} = \bigl\{{\ts\frac 12}\bigr\}^\pm_{1} \,,
 \quad\quad\ \label{k1a} 
 \sum_{m=0}^\infty \frac{(\pm 1)^m q^{\frac 12 m^2-\frac 12 m }}
 {(q)_m} = (1\pm 1)\, \{1\}^+_1 \,, \\
 && \!\!\!\!\!\!\!\!\!\!\!\!
 \sum_{m=0}^\infty \frac{(\pm 1)^{m}\,q^{\frac 14 m^2+(1-\frac n2)m }}
 {(q)_m} \= \frac{ \bigl\{ \frac{5-2n}{4};\frac{5+2n}{4} 
 \bigr\}^\pm_{\frac 52} }{ \{ \frac 52\}^+_{\frac 52}\{3-n;2+n\}^-_5 } 
 \,,\quad n=1,2 \,, \label{k1b} \\
 && \!\!\!\!\!\!\!\!\!\!\!\!
 \sum\limits_{\vec{m}=\vec{0}}^\infty 
 \frac{(\pm 1)^{m_2} q^{ m_1^2 + \frac 34 m_2 + m_1 m_2+ 
 \delta_{1n}m_1 +(1-\frac n2)m_2 }}{ (q)_{m_{1}} (q)_{m_{2}}} =
 \frac{ \bigl\{\frac{7-2n}{4};\frac{7+2n}{4}\bigr\}^\pm_{\frac 72} }
 {\bigl\{\frac 72\bigr\}^+_{\frac 72}
 \!\! \prod\limits_{j=1 \atop j\neq n}^3  \{j;7-j\}^-_7 
 } \,,\quad 1\leq n \leq 3 \,, \label{k2a} \\
 && \!\!\!\!\!\!\!\!\!\!\!\!
 \sum\limits_{\vec{m}=\vec{0}}^\infty 
 \frac{(\pm 1)^{m_2} q^{ m_1^2 + \frac 12 m_2 + m_1 m_2+ 
 \delta_{1n}m_1 +(\frac 32 -\frac n2)m_2 }}{ (q)_{m_{1}} (q)_{m_{2}}} 
 =\frac{ \bigl\{2-\frac{n}{2};2+\frac{n}{2}\bigr\}^\pm_{4} }
 { \prod\limits_{j=1 \atop j\neq n}^3 
 \{j;8-j\}^-_8 } \,,\quad 1\leq n \leq 3 \,. \label{k2b} 
\ea
Here we simplified the product sides exploiting the Euler identity 
$\{x\}_{x}^+ \{x\}_{2x}^- =1$ and other transformations (see \cite{BF2}).
Eqs.~(\ref{k1a}) are well known (see e.g.~\cite{Sl}), eqs.~(\ref{k1b})
were presented in \cite{BF2}. It should be remarked that in some cases
the combinations on the l.h.s.~of (\ref{M31})-(\ref{M32}) belong both
to (\ref{f3}) and (\ref{f4}). In this case the product side acquires 
more compact form \cite{BF2}:
\be{f34}
 \chi_{m,n}^{3m,4n}(q) \pm \chi_{m,3n}^{3m,4n}(q)
 \= \frac{q^{\,\frac{nm-2}{48}}}{ \{1\}^-_1 } 
  \{nm\}^-_{nm} \{ \frac{nm}{2} \}^\pm_{nm} \, .
\ee
Therefore we obtain 
\ba
 \sum\limits_{\vec{m}=\vec{0}}^\infty 
 \frac{(\pm 1)^{m_{4t+1}} q^{\vec{m}^{\,t} A_{4t+1} \vec{m} + 
 \vec{m}\cdot\vec{B}^{12t+4}_{3t+1}}}{(q)_{m_{1}} \ldots (q)_{m_{4t+1}}}
 \=\frac{ \{3t+1\}^-_{3t+1} \{\frac{3t+1}{2}\}^\pm_{3t+1} }{ 
 \{1\}^-_1}\,, \label{rr3} \\
 \sum\limits_{\vec{m}=\vec{0}}^\infty 
 \frac{(\pm 1)^{m_{4t+2}} q^{\vec{m}^{\,t} {\tilde A}_{4t+2} \vec{m} + 
 \vec{m}\cdot\vec{B}^{12t+8}_{3t+2}}}{(q)_{m_{1}} \ldots (q)_{m_{4t+2}}}
 \=\frac{ \{3t+2\}^-_{3t+2} \{\frac{3t+2}{2}\}^\pm_{3t+2} }{ \{1\}^-_1}
 \,, \label{rr4}
\ea
where $t=0,1,2,\ldots$, and $A$, ${\tilde A}$ and $\vec{B}$ are 
defined in (\ref{A3a})-(\ref{A3c}) and (\ref{B3})-(\ref{B4}). 
For instance, (\ref{rr4}) yields for $t=0$ 
\be{38}
 \sum_{\vec{m}=\vec{0}}^\infty \frac{q^{m_1^2+ \frac 12 m_2^2+
 m_1m_2+ \frac 12 m_2}}{ (q)_{m_1} (q)_{m_2}} =
 \{1\}^+_1 \{1\}^+_2 \,, \quad
 \sum_{\vec{m}=\vec{0}}^\infty \frac{(-1)^{m_2} q^{m_1^2+ 
 \frac 12 m_2^2+ m_1m_2+ \frac 12 m_2}}{ (q)_{m_1} (q)_{m_2}} = 1 \, .
\ee
The last equality is due to the fact that
$\chi^{3,8}_{1,2}(q)-\chi^{3,8}_{1,6}(q)=1$ (see \cite{BF2,CIZ}).

For the fermionic sums (\ref{M33})-(\ref{M34}) the product side 
also simplifies since these characters belong both to (\ref{2s}) and 
(\ref{3s}). Namely, we obtain for $t=1,2,3\ldots$
\be{rr5} 
 \sum\limits_{\vec{m}=\vec{0}}^\infty \frac{ q^{\vec{m}^{\,t} A_{2t-1} 
 \vec{m} + \vec{m}\cdot\vec{C}^{2t-1}}}{(q)_{m_{1}} \ldots (q)_{m_{2t-1}}} 
 \= \frac{ \{3t-1\}^-_{3t-1} }{ \{1\}^-_1 } \,,\quad\quad
 \sum\limits_{\vec{m}=\vec{0}}^\infty \frac{ q^{\vec{m}^{\,t} {\tilde 
 A}_{2t} \vec{m} + \vec{m}\cdot\vec{C}^{2t} }} {(q)_{m_{1}} \ldots 
 (q)_{m_{2t}}} \= \frac{ \{3t+1\}^-_{3t+1} }{ \{1\}^-_1 } \,,
\ee
where $A$, ${\tilde A}$ and $\vec{C}$ are defined in 
(\ref{A3a})-(\ref{A3c}) and (\ref{M35}). For instance, for $t=1$ we get
\be{rr6}
  \sum_{m=0}^\infty \frac{ q^{\frac 12 m^2+\frac 12 m} }{(q)_m} \=
  \{1\}^+_1 \,, \quad\quad 
 \sum_{\vec{m}=\vec{0}}^\infty \frac{q^{m_1^2+ \frac 12 m_2^2+
 m_1m_2+ m_1+\frac 12 m_2}}{ (q)_{m_1} (q)_{m_2}} =
 \{1\}^+_1 \{2\}^+_2 \,.
\ee

To complete the discussion of the ${\cal M}(3,t)$ case, we recall that
since each of eqs.~(\ref{M31})-(\ref{M32}) comprises two equalities,
we can express the involved characters as sums without alternation of the
sign but with summation over $m_k$ restricted to odd or even numbers. 
Then, according to (\ref{3s}), each character obtained in this way
admits also the product representation. Thus, we get for $t=3k{+}1$,\,
$1{\leq} n {\leq}  k{+}1$
\ba 
 && \sum\limits_{\vec{m}=\vec{0} \atop m_k{\rm -even}}^\infty \!\!
 \frac{q^{\vec{m}^{\,t} A_k \vec{m} + \vec{m}\cdot\vec{B}^{t}_{n}}}{ 
 (q)_{m_{1}} \ldots  (q)_{m_{k}}} \= \frac{1}{\{1\}^-_1}  \{n;2t-
 n;2t\}_{2t}^- \{2t-2n;2t+2n\}_{4t}^- \,, \\ 
 && \sum\limits_{\vec{m}=\vec{0} \atop m_k{\rm -odd}}^\infty \!\!
 \frac{q^{\vec{m}^{\,t} A_k \vec{m}+\vec{m}\cdot\vec{B}^{t}_{n} }}
 {(q)_{m_{1}} \ldots (q)_{m_{k}}} \=  \frac{1}{\{1\}^-_1}
 \{t-n;t+n;2t\}_{2t}^- \{2n;4t-2n\}_{4t}^- \,,
\ea
and analogous formulae for $t=3k{+}2$ if $A_k$ is replaced by 
$\tilde{A}_k$. 
\vspace*{3pt}

\noindent
\underline{${\cal M}(4,5)$:}\ From (\ref{M45b}) and (\ref{2s}) we
obtain for $n=1,2$
\be{RR5}
 \sum_{\vec{m}=\vec{0} }^\infty  \frac{ q^{2m_1^2+\frac 12 m_2^2+
 m_1m_2+(4-2n)m_1+ \frac 12 m_2} } { (q)_{m_1} (q)_{m_2} } \=
 \frac{\{5\}_5^+}{ \{n;5-n\}_{5}^{-} \{5-2n;5+2n\}_{10}^{-} } \,.
\ee

\noindent
\underline{${\cal M}(5,6)$:}\ Combining (\ref{M56b})-(\ref{M56e}) with 
(\ref{f3}) and (\ref{f6}), we obtain for $n=1,2$
\ba 
&& \!\!\!\!\!\!\!\!\!\!
 \sum_{\vec{m}=\vec{0}}^\infty \frac{ (\pm 1)^{m_2} q^{\frac 12 (m_1^2
 +m_2^2) +m_1m_2+ \frac 12 m_1+(2-n)m_2} }{ (q)_{m_1} (q)_{m_2}}
 \= \frac{1}{ \{n;5-n\}_{5}^{-}  \bigl\{\frac 52-n;\frac 52 +n 
 \bigr\}_{5}^{\mp} } =  \label{RR6} \\
&& = \sum_{\vec{m}=\vec{0}}^\infty\frac{ (\pm 1)^{m_3} q^{ 
 m_1^2 +2m_2^2+ m_3^2 + 2m_1m_2 +m_1m_3 +2m_2m_3 +(2-n)m_2 +
(\frac 32-n)m_3} }{ (q)_{m_1} (q)_{m_2} (q)_{m_3}} \,, \nn\\
&& \!\!\!\!\!\!\!\!\!\!
 \sum_{\vec{m}=\vec{0}}^\infty  \frac{ (-1)^{m_2} q^{\frac 12 (m_1^2 +
 m_2^2) +m_1m_2+ (n-1)(m_1+m_2)} }{ (q)_{m_1} (q)_{m_2} } = 
\frac{1}{ \{2n;10-2n\}_{10}^{-}  } = \label{RR7} \\
&& = \sum_{\vec{m}=\vec{0}}^\infty \frac{ (-1)^{m_3} q^{ 
 m_1^2 +2m_2^2+ m_3^2 + 2m_1m_2 +m_1m_3 +2m_2m_3 +
 (n-1)(m_1+2m_2 +m_3)} }{ (q)_{m_1} (q)_{m_2} (q)_{m_3}} \,. \nn
\ea

\noindent
\underline{${\cal M}(6,7)$:}\ The fermionic sums on the r.h.s.~of (\ref{M67a}) 
and (\ref{M67c}) are equal, respectively, to
\be{RR8} 
 \frac{q^{\eta^{6,7}_{2,n}}}{ \{1;4-n;3+n;6\}_{7}^- 
 \{\frac 72 -n;\frac 72 +n \}^\mp_7 } \,, \quad\quad
 \frac{q^{\eta^{6,7}_{1,n}}}{ \{d_n;7-d_n\}_{7}^{-} 
 \{2n;14-2n\}_{14}^{-} } \,, 
\ee
where $d_1=3$, $d_2=1$, $d_3=2$.

\section{Dilogarithm identities.} 

If a $q$-series $\chi(q)$ (not necessarily identified in terms of characters) 
admits both fermionic representation (\ref{F1}) and
product representation (\ref{fact}), it implies not only existence of a 
Rogers-Ramanujan type identity but also leads to a certain identity 
involving the dilogarithm function, 
$L(x)=\sum_{n=1}^{\infty} \frac{x^n}{n^2} + \frac 12 \ln{x}\ln(1-x)$.
Indeed, the product side allows us to find easily the number $c_{\rm eff}$
that governs the asymptotics of $\chi(q)$ in the $q\rightarrow 1$ limit
(see e.g.~\cite{BF2,BF1}). On the other hand, the same number can be
obtained from the fermionic sum by the saddle point analysis 
(see e.g.~\cite{KKMM,NRT}). 
Equivalence of the two expressions for $c_{\rm eff}$
is  typically a nontrivial identity. Of course, if it is known a-priory
that $\chi(q)$ is a character, then its fermionic form alone leads to a 
dilogarithm identity since $c_{\rm eff}$ (effective central charge) is
fixed by the properties of $\chi(q)$ with respect to the modular 
transformations. Namely, let $q=e^{2\pi i \tau}$ and 
$\hat{q}=e^{-2\pi i /\tau}$, then for the minimal Virasoro model 
${\cal M}(s,t)$ we have 
$\chi^{s,t}_{n,m}(q)\sim \hat{q}^{-\frac{c_{\rm eff}(s,t)}{24}}$ as 
$q\rightarrow 1$, where
\be{ce}
  c_{\rm eff}(s,t) \= c(s,t) - 24 h^\prime \= 1- \frac{6}{st}  \,.
\ee
Here $h^\prime$ denotes the lowest conformal weight in the model.
Furthermore, as it was shown in \cite{BF2}, a difference of characters
of the type (\ref{pm}) for all minimal models but ${\cal M}(2,t)$
has the asymptotics $\hat{q}^{-\frac{\tilde{c}(s,t)}{24}}$ when 
$q\rightarrow 1$. Here $\tilde{c}$ (secondary effective central charge)
is given by
\be{ct}
  \tilde{c}(s,t) \= c(s,t) - 24 h^{\prime\prime} \= 1- \frac{24}{st} \,,
\ee
where $h^{\prime\prime}$ stands for the second lowest conformal weight
in the model.

For our purposes we need to consider a slightly generalized version of
(\ref{F1}):
\be{F2}
\chi(q) \= q^{\rm{const}} \sum\limits_{\vec{m}} 
 \frac{q^{\vec{m}^{\,t} A\vec{m}+ \vec{m}\cdot \vec{B}
 }}{(q^{b_1})_{m_{1}}\ldots (q^{b_r})_{m_{r}}} \,,
\ee
where $b_i$ are some positive numbers.
Modifying properly the standard saddle point analysis of a fermionic
sum (see \cite{KKMM,NRT} for the case $b_i=1$ and \cite{BF1} for 
$b_i=b\neq 1$), we find that (\ref{F2}) has the asymptotics
$\hat{q}^{-\frac{c_{\rm eff}}{24}}$ as $q\rightarrow 1$ with 
\be{cb}
 c_{\rm eff} \= \frac{6}{\pi^2} \sum_{i=1}^r \frac{1}{b_i} L(x_i) \,.
\ee
Here the set of numbers $0<x_i<1$ satisfies the following equations
\be{xb}
 x_i \= \prod_{j=1}^r (1-x_j)^{\frac{1}{b_j}(A_{ij}+A_{ji})}  
 \,, \quad\ i=1,\ldots,r \,.
\ee

Let us define $(\widehat{A})_{ij}=\frac{1}{2b_j}(A_{ij}+A_{ji})$. 
If matrix $\widehat{A}$ is invertible, it is convenient to introduce 
$I=2-(\widehat{A})^{-1}$ (generalized incidence matrix) and make
the substitution $x_i=1/\mu^2_i$. Then (\ref{cb}) and (\ref{xb})
turn into
\be{cm}
 c_{\rm eff} \=\frac{6}{\pi^2} \sum_{i=1}^r \frac{1}{b_i} 
 L(\frac{1}{\mu^2_i}) \,,\quad\quad  
 \mu_i^2 \= 1 + \prod_{j=1}^r (\mu_j)^{I_{ij}} \,.
\ee

As we have seen above, certain differences of characters of the type
(\ref{pm}) admit the fermionic form with alternated summation over the 
last variable,
\be{cha}
 \chi(q) \= q^{\rm{const}}
\sum\limits_{\vec{m}} \frac{(-1)^{m_r} q^{\vec{m}^{\,t}A\vec{m}+
\vec{m}\cdot \vec{B} }}{(q)_{m_{1}} \ldots (q)_{m_{r}}} \, .  
\ee
Let us find equations describing the $q\rightarrow 1$ limit of such
series. To this end we notice that 
\be{tr}
 \frac{1}{(q)_m} \=(-1)^m q^{-\,\frac 12 m(m+1)}\frac{1}{(q^{-1})_m } \,.
\ee
Therefore we can rewrite (\ref{cha}) as follows
\be{cha2}
 \chi(q) \= q^{\rm{const}} \sum\limits_{\vec{m}} 
 \frac{q^{\vec{m}^{\,t} A^{\prime} \vec{m}+ \vec{m}\cdot \vec{B}^{\prime} 
  }}{(q)_{m_{1}} \ldots (q)_{m_{r-1}}(q^{-1})_{m_{r}}} \, ,  
\ee
where 
\be{Ap}
 (A^{\prime})_{ij} \= A_{ij}- {\ts \frac 12} \delta_{ir}\delta_{jr} 
 \,,\quad\quad   
 (\vec{B}^{\prime})_j \= (\vec{B})_j- {\ts \frac 12} \delta_{jr} \,.
\ee 
Eq.~(\ref{cha2}) is a particular case of (\ref{F2}) with 
$b_1=\ldots=b_{r-1}=-b_r=1$ and thus we can apply 
eqs.~(\ref{cb})-(\ref{xb}). We conclude that (\ref{cha}) has the 
asymptotics $\hat{q}^{-\frac{\tilde{c}}{24}}$ as $q\rightarrow 1$ with
\be{ca}
 \tilde{c} \=\frac{6}{\pi^2}
 \Bigr(\sum_{i=1}^{r-1} L(y_i) - L(y_r) \Bigl) \, ,
\ee
where the set of numbers $0< y_i <1$ satisfies the following equations
\be{ya}
  y_i \= \prod_{j=1}^r (1-y_j)^{ (-1)^{\delta_{jr}} 
 (A^{\prime}_{ij}+A^{\prime}_{ji})} \,, \quad i=1,\ldots,r \, .
\ee

It is again convenient to introduce 
$I^\prime=2-({\widehat{A}}^\prime)^{-1}$, where
$({\widehat{A}}^\prime)_{ij}=\frac 12 (-1)^{\delta_{jr}}(A^\prime_{ij}+
A^\prime_{ji})$.  Then, making the substitution $y_i=1/\nu^2_i$, we 
transform (\ref{ca})-(\ref{ya}) to
\be{cn}
 \tilde{c} \=\frac{6}{\pi^2} \Bigr(\sum_{i=1}^{r-1} 
 L(\frac{1}{\nu^2_i})- L(\frac{1}{\nu^2_r}) \Bigl) \,,\quad\quad
 \nu_i^2 \= 1 + \prod_{j=1}^r (\nu_j)^{ I^\prime_{ij}} \,.
\ee

Let us remark that performing the following change of variables in 
(\ref{ca})-(\ref{ya}): $z_i=y_i$, $i<r$ and $z_r=\frac{y_r}{y_r-1}$,
we can transform these equations to the form almost coinciding with 
(\ref{cb})-(\ref{xb}) (for $b_i=1$) and involving the initial matrix $A$:
\be{za}
 \tilde{c} \=\frac{6}{\pi^2} \sum_{i=1}^{r} L(z_i) \,, \quad\quad
 (-1)^{\delta_{ir}} z_i \= \prod_{j=1}^r (1-z_j)^{A_{ij}+A_{ji}}  
 \,, \quad\ i=1,\ldots,r \,.
\ee
In contrast with (\ref{xb}), now $z_r<0$. Deriving (\ref{za}) we used 
the definition \cite{Lew,Kir}: $L(x)=L(\frac{1}{1-x})-L(1)$ for $x<0$, 
and the property $L(x)=-L(\frac{x}{x-1})$ for $x<1$.

Generalization of (\ref{ca})-(\ref{za}) for the case of a fermionic
sum involving alternated summation over several variables is obvious.
Now let us list dilogarithm identities that follow from the formulae 
(\ref{cm}), (\ref{cn}) and (\ref{ce})-(\ref{ct}) for the (combinations 
of) characters considered in the previous sections.

\noindent
\underline{${\cal M}(3,3k+1)$ and ${\cal M}(3,3k+2)$}: \ 
The explicit expressions (\ref{A3a})-(\ref{A3c}) for the matrices $A_k$
and ${\tilde A}_k$ allow us to compute the corresponding matrices $I_k$
and $\tilde{I}_k$
\ba  
 && \bigl(I_k\bigr)_{ij} \= \delta_{i,j+1} + \delta_{i+1,j} + 
 {\ts \frac 12} \delta_{i,k-1} \delta_{j,k-1} \,, \label{I1} \\
 && \bigl(\tilde{I}_k\bigr)_{ij} \= \delta_{i,j+1} + \delta_{i+1,j} + 
 \delta_{i,k} \delta_{j,k-1} + \delta_{i,k-1} \delta_{j,k} -
 2 \delta_{i,k} \delta_{j,k} \,. \label{I2}
\ea
These generalized incidence matrices differ from those of the Lie algebra
$A_k$ only by a few entries in the lower-right corner. This hints on a 
possibility to solve the corresponding sets of equations (\ref{cm}) in a 
uniform manner, similar to that known for the ${\cal M}(2,t)$ models 
\cite{NRT,KM}. Indeed, we find the following solutions of (\ref{cm}) 
(they can be verified by a straightforward substitution):
\be{m1}
 \mu_i \= \frac{\sin\frac{(i+1)\pi}{3k+1}}{\sin\frac{\pi}{3k+1}} \,,
 \quad \ 1\leq i \leq k-1 \,; \quad\quad\
 \mu_k^2 \= 1+ \frac{\sin\frac{k\pi}{3k+1}}{\sin\frac{\pi}{3k+1}} 
\ee
for $I_k$ given by (\ref{I1}) (that is in the case of ${\cal M}(3,3k+1)$ 
model) and
\be{m2}
 \mu_i \= \frac{\sin\frac{(i+1)\pi}{3k+2}}{\sin\frac{\pi}{3k+2}} \,,
 \quad \ 1\leq i \leq k-1 \,; \quad\quad\
 \mu_k^2 \= \frac{\sin\frac{(k+1)\pi}{3k+2}}{\sin\frac{\pi}{3k+2}} 
\ee
for $\tilde{I}_k$ given by (\ref{I2}) (${\cal M}(3,3k+2)$ model). 
Combining these results with (\ref{ce}), we derive the following identities
\ba
 \label{di1} && \sum_{i=1}^{k-1} L\Bigl(\frac{\sin^2\frac{\pi}{3k+1}}  
 {\sin^2\frac{(i+1)\pi}{3k+1}}\Bigr) + L\Bigl(\frac{\sin\frac{\pi}{3k+1}}
 {\sin\frac{\pi}{3k+1}+\sin\frac{k\pi}{3k+1}}\Bigr) 
 \= \frac{\pi^2}{6} \,\frac{3k-1}{3k+1} \,,\\  
 \label{di2} && \sum_{i=1}^{k-1} L\Bigl(\frac{\sin^2\frac{\pi}{3k+2}}
 {\sin^2\frac{(i+1)\pi}{3k+2}}\Bigl) + L\Bigr(\frac{\sin\frac{\pi}{3k+2}}
 {\sin\frac{(k+1)\pi}{3k+2}}\Bigl) \=\frac{\pi^2}{6} \,\frac{3k}{3k+2} \,.
\ea
Let us remark that although these identities were derived here exploiting 
the modular properties of characters and the saddle point analysis, they 
resemble the ``general $A_1$-type" dilogarithm identities \cite{Kir} 
and probably can be proved in more direct way based on the functional 
relations for the dilogarithm. For instance, eq.~(\ref{di2}) for $k=2$
yields the equality $\bigl(L(1-\frac{1}{\sqrt{2}})+
 L(\sqrt{2}-1)\bigr)=\frac{\pi^2}{8}$. It can be proved with the help
of the Abel duplication formula \cite{Lew,Kir}. It should be mentioned
that eq.~(\ref{di1}) for $k$ odd was encountered in \cite{Ahn} in the
context of the thermodynamic Bethe ansatz.

Next we consider the differences of characters in (\ref{M31})-(\ref{M32}).
First, we compute the matrices $I^\prime_k$ and $\tilde{I}^\prime_k$. It 
turns out that for ${\cal M}(3,3k+1)$\, $\det({\widehat{A}_k}^\prime)=0$
that is $I^\prime_k$ does not exist and we have to solve the equations 
(\ref{ya}). For ${\cal M}(3,3k+2)$ the matrix $\tilde{I}^\prime_k$ exists 
and is given by
\be{I3}
 \bigl(\tilde{I}_k^\prime\bigr)_{ij} \= \delta_{i,j+1} + \delta_{i+1,j}
 + \delta_{i,k} \delta_{j,k-1} - 3\delta_{i,k-1} \delta_{j,k} +
 2\delta_{i,k-1} \delta_{j,k-1} -2\delta_{i,k} \delta_{j,k} \,.
\ee
We obtain the following solution of (\ref{ya}) for ${\cal M}(3,3k+1)$
(written in terms of $\nu_i=1/\sqrt{y_i}$ for the sake of uniformness)
\be{m3}
 \nu_i \= \frac{\sin\frac{(2i+2)\pi}{3k+1}}{\sin\frac{2\pi}{3k+1}} \,,
 \quad \ 1\leq i \leq k-1 \,; \quad\quad\
 \nu_k^2 \= \frac{\sin\frac{2k\pi}{3k+1}}{\sin\frac{2\pi}{3k+1}} \,.
\ee
and of (\ref{cn}) for ${\cal M}(3,3k+2)$:
\be{m4}
 \nu_i \= \frac{\sin\frac{(2i+2)\pi}{3k+2}}{\sin\frac{2\pi}{3k+2}} \,,
 \quad \ 1\leq i \leq k-1 \,; \quad\quad\
 \nu_k^2 \= 1+\frac{\sin\frac{(2k+2)\pi}{3k+2}}{\sin\frac{2\pi}{3k+2}}\,. 
\ee
Combining these results with (\ref{ct}), we obtain the following 
dilogarithm identities
\ba
 \label{di3} &&  \sum_{i=1}^{k-1} L\Bigl(\frac{\sin^2\frac{2\pi}{3k+1}}  
 {\sin^2\frac{(2i+2)\pi}{3k+1}}\Bigr) - L\Bigl(\frac{\sin\frac{2\pi}{3k+1}}
 {\sin\frac{2k\pi}{3k+1}}\Bigr) \=\frac{\pi^2}{6}\,\frac{3k-7}{3k+1} \,,\\  
 \label{di4} && \sum_{i=1}^{k-1} L\Bigl(\frac{\sin^2\frac{2\pi}{3k+2}}
 {\sin^2\frac{(2i+2)\pi}{3k+2}}\Bigl) -L\Bigr(\frac{\sin\frac{2\pi}{3k+2}}
 {\sin\frac{2\pi}{3k+2}+\sin\frac{(2k+2)\pi}{3k+2}} \Bigl) 
  \= \frac{\pi^2}{6} \,\frac{3k-6}{3k+2} \, .
\ea

\noindent
\underline{${\cal M}(4,5)$}:\ Eq.~(\ref{xb}) for $A$ given by (\ref{M45a}) 
can be reduced to a bi-quadratic equation. Solving it we obtain 
$x_1=1-\sqrt{\rho}$, $x_2=1-\frac{1}{1+\sqrt{\rho}}$, where 
$\rho:=\frac{\sqrt{5}-1}{2}$. According to (\ref{ce}) and (\ref{cb})
this leads to the identity
\be{di5}
 {\ts L\Bigl(1-\sqrt{\frac{\sqrt{5}-1}{2}}\Bigr)+
 L\Bigl(1- \frac{\sqrt{2}}{\sqrt{2}+\sqrt{\sqrt{5}-1}}\Bigr) } 
 \= \frac{7}{10} \frac{\pi^2}{6} \,.
\ee
Although the summation in the fermionic representation (\ref{M45b}) is 
not alternated, we can nevertheless introduce matrix $A^\prime$ according
to (\ref{Ap}) and solve eqs.~(\ref{ca})-(\ref{ya}). The solution is
$y_1=1-\rho$, $y_2=\rho$ and thus we get 
$\tilde{c} = \frac{6}{\pi^2} \bigl(L\bigl(1-\rho) - L\bigl(\rho)\bigr) 
 = -\frac 15 $. This identity is rather trivial mathematically (since it 
is well known that $L(\rho)=\frac{\pi^2}{10}$ and 
$L(1-\rho)=\frac{\pi^2}{15}$) but it is remarkable that the value 
of $\tilde{c}$ is in agreement with (\ref{ct}). 

\noindent
\underline{${\cal M}(5,6)$}:\ It is easy to see that for this model any
matrix of the form
\be{Aa}
 A \=\left(\ar{cc} 1-\alpha & \alpha \\ \alpha & 1-\alpha \er\right)\,,
 \qquad \alpha \leq 1/2 \,,
\ee 
yields $x_1=x_2=1-\rho$ and gives the correct central charge:
$c_{\rm eff}=\frac{6}{\pi^2} 2L(1-\rho)=\frac 45$. Besides our example
(\ref{M56a}) corresponding to $\alpha=\frac 12$ the case of
$\alpha=\frac 13$ is known \cite{KKMM}. However, matrix $A^\prime$ 
constructed from (\ref{Aa}) according to (\ref{Ap}) leads to the correct 
value of $\tilde{c}$ only for $\alpha=\frac 12$. Namely, in this case we
get $y_1=\rho$, $y_2=1-\rho$ and $\tilde{c} =\frac{6}{\pi^2} \bigl(L(\rho)
 - L(1-\rho)\bigr) =\frac 15$. It would be interesting to see if there
are fermionic representations for characters of ${\cal M}(5,6)$
corresponding to other values of $\alpha$. Below we will show that
$\alpha=0$ appears not in ${\cal M}(5,6)$ but in closely related 
${\cal M}(3,10)$ model.

\section{Further fermionic representations and identities.}

So far we considered only ``irreducible" fermionic representations
of characters, i.e.~those that are not decomposable into a product of
other fermionic sums. However, there are many ways to construct
``reducible" fermionic representations. One of them was discussed in
\cite{BF2,BF1} and based on the fact that any factorizable character
can be brought to the form (\ref{F2}) (typically with $b_i\neq 1$)
with the help of the following formulae 
\be{PF}
 \{x\}^\pm_y \= \sum_{m=0}^\infty \frac{(\pm 1)^m q^{\frac y2 (m^2-m) 
  +mx}}{(q^y)_m} \,, \quad\quad\ \frac{1}{\{x\}^\pm_y} \= 
 \sum_{m=0}^\infty \frac{ (\mp 1)^m q^{mx}}{(q^y)_m} \,.
\ee

Another possibility is to use relations expressing a character as a 
product of other characters for which fermionic representation is known.
To derive and prove such relations it is often convenient to exploit
the factorized forms of characters. For instance, with the help of
(\ref{2s}), (\ref{3s}) and (\ref{f3}) it is straightforward to check
the following identities (see \cite{BF2} for a similar derivation)
\ba
 && \!\!\!\! \chi^{2n,3m}_{n,m}(q) \chi^{3n,10m}_{n,5m}(q) \=
 \chi^{2n,5m}_{n,m}(q) \,\chi^{2n,5m}_{n,2m}(q) \,, \label{cc1}\\
 && \!\!\!\! \chi^{2n,3m}_{n,m}(q) \, \Bigl( \chi^{3n,10m}_{n,m(2k-1)}(q)
 + \chi^{3n,10m}_{n,m(11-2k)}(q) \Bigr) \=  \chi^{2n,5m}_{n,mk}(q) 
 \,\chi^{2n,5m}_{n,mk}(q) \,, \quad k=1,2\,. \label{cc2}
\ea
Choosing here $n=m=1$ (recall that $\chi^{2,3}_{1,1}(q)=1$) and using
the sum side of the Rogers-Ramanujan identities (\ref{RR}), we obtain
the following  formulae
\ba
 && \!\!\!\!\!\!\!\!\!\!\!\!\!\!\!
 \chi^{3,10}_{1,5}(q) =\chi^{2,5}_{1,1}(q) \chi^{2,5}_{1,2}(q) =  
 q^{\frac 16} \sum_{\vec{m}=\vec{0}}^\infty 
 \frac{ q^{m_1^2 +m_2^2 + m_1}}{(q)_{m_1} (q)_{m_2}} \,, \label{cc3} \\
 && \!\!\!\!\!\!\!\!\!\!\!\!\!\!\!
 \chi^{3,10}_{1,2k-1}(q)+\chi^{3,10}_{1,11-2k}(q) = \chi^{2,5}_{1,k}(q) 
 \chi^{2,5}_{1,k}(q) =  q^{\frac{23-12k}{30}} 
 \sum_{\vec{m}=\vec{0}}^\infty \frac{q^{m_1^2 +m_2^2 + (4-2k)m_1}}
 {(q)_{m_1} (q)_{m_2}} , \label{cc4}
\ea
where $k=1,2$.
These reducible fermionic representations correspond to the choice 
$\alpha=0$ for the matrix $A$ given by (\ref{Aa}).

Using in the similar way the following identities for $k=1,2$
(they were found in \cite{Tao} and generalized to the form similar to 
(\ref{cc1})-(\ref{cc2}) in \cite{BF2})
\ba
 && \label{cc5} \chi_{2,k}^{4,5}(q) \= \chi_{1,2}^{3,4}(q) \,
 \Bigl(\chi_{1,(3-k)}^{6,5}(q) - \chi_{1,(2+k)}^{6,5}(q) \Bigr) \,, \\
 && \label{cc6} \chi_{1,k}^{4,5}(q) \pm \chi_{3,k}^{4,5}(q) \=
 \Bigl(\chi_{1,1}^{3,4}(q) \pm \chi_{1,3}^{3,4}(q) \Bigr) \,
 \Bigl(\chi_{2,(3-k)}^{6,5}(q) \mp \chi_{2,(2+k)}^{6,5}(q) \Bigr) 
\ea
and employing any of the fermionic representations for ${\cal M}(3,4)$
and ${\cal M}(5,6)$ discussed above, we get different reducible fermionic
representations for ${\cal M}(4,5)$. For instance, substituting
(\ref{34}) and (\ref{M56c}) into (\ref{cc5}) we obtain
\be{cc7}
 \chi_{2,k}^{4,5}(q) \=q^{\eta^{4,5}_{2,k}} \sum_{\vec{m}=\vec{0}}^\infty
 \frac{ (-1)^{m_3} q^{\frac 12 (m_1^2 +m_2^2 +m_3^2)+m_2m_3+\frac 12 m_1 
 +(2-k)(m_2+m_3) } }{ (q)_{m_1} (q)_{m_2} (q)_{m_3} } \,,\quad k=1,2 \,.
\ee 

One more possibility to extend the list of fermionic representations
is to consider relations between characters with rescaled argument $q$.
For instance, we have \cite{BF2}
\be{q1}
 \chi_{n,2}^{5,6}(q) + \chi_{n,4}^{5,6}(q) \= 
 \chi_{1,n}^{2,5}(q^{\frac 12})\,, \quad\quad
 \chi_{n,1}^{5,6}(q) - \chi_{n,5}^{5,6}(q) \= 
 \chi_{1,3-n}^{2,5}(q^2)\,, \quad n=1,2\,. 
\ee
Together with (\ref{RR}) this gives yet another fermionic representation
for ${\cal M}(5,6)$
\be{q2}
 \chi_{n,2}^{5,6}(q) + \chi_{n,4}^{5,6}(q) \=
 \sum_{m=0}^\infty \frac{ q^{\frac 12 m^2 + (1-\frac n2)m}}
 {(q^{\frac 12})_m}\,, \quad\quad 
 \chi_{n,1}^{5,6}(q) - \chi_{n,5}^{5,6}(q) \=   \sum_{m=0}^\infty 
 \frac{q^{2 m^2+(2n-2)m}}{(q^2)_m} \,.
\ee
On the other hand, reading eqs.~(\ref{q1}) from the right to the left
and using (\ref{M56b})-(\ref{M56c}), we obtain alternative sum sides
for the Rogers-Ramanujan identities (\ref{RR}):
\be{q3}
 \sum_{m=0}^\infty \frac{q^{m^2+(2-n)m}}{(q)_m} \=
 \sum_{\vec{m}=\vec{0}}^\infty \frac{ q^{(m_1 +m_2)^2 + 
 m_1+(4-2n)m_2} }{ (q^2)_{m_1} (q^2)_{m_2}}  \=
 \sum_{\vec{m}=\vec{0}}^\infty  \frac{ (-1)^{m_2} 
 q^{\frac 14 (m_1 + m_2)^2 + (1-\frac n2 )(m_1+m_2)} }{ 
 (q^{\frac 12})_{m_1} (q^{\frac 12})_{m_2} } \,, 
\ee
where $n=1,2$. This sequence of identities can be continued further 
employing the fermionic representations (\ref{M56d})-(\ref{M56e}) and 
also those found in \cite{KKMM} (corresponding to (\ref{Aa}) with
$\alpha=\frac 13$).
 
Another set of relations observed in \cite{BF2}
\be{q4}
 \chi_{2,n}^{6,7}(q^2) + \chi_{4,n}^{6,7}(q^2) \= 
 \chi_{1,d_n}^{6,7}(q) - \chi_{5,d_n}^{6,7}(q) \,,
\ee
where $d_1=3$, $d_2=1$, $d_3=2$, together with 
eqs.~(\ref{M67a})-(\ref{M67c}) can be used to get fermionic 
representations of the type (\ref{F2}) for the ${\cal M}(6,7)$ model.

It is also possible to combine rescaling of $q$ and constructing
of reducible fermionic sums. For example, let $p$ be a prime number such
that $\langle p,t \rangle =1$. Then, using (\ref{char}), it is easy
to derive the following relation
\be{qp}
  \chi_{1,p}^{3,2p}(q) \, \chi_{n,m}^{s,t}(q^p) \=
  \chi_{pn,m}^{ps,t}(q) \,.
\ee
Here $\chi_{1,p}^{3,2p}(q)$ should be replaced with $\chi_{1,3}^{2,9}(q)$
if $p=3$. If $p$ is not a prime number, generalization of (\ref{qp}) 
can be achieved by decomposing $p$ into proper factors. Now choosing 
$p=2$, $t=5$ and $s=2$ or $s=3$ in (\ref{qp}), we get for $k=1,2$
\be{qq2}  
 \chi_{2,k}^{4,5}(q) =\chi_{1,k}^{2,5}(q^2) \chi_{1,2}^{3,4}(q) \,,
 \quad\quad    \chi_{k,2}^{5,6}(q) \pm \chi_{k,4}^{5,6}(q)
 = \Bigl( \chi_{1,k}^{3,5}(q^2) \pm \chi_{1,5-k}^{3,5}(q^2) \Bigr) 
 \chi_{1,2}^{3,4}(q) \,.
\ee
Substituting here (\ref{2t}), (\ref{34}) and (\ref{k1b}), we obtain more 
fermionic representations of the type (\ref{F2}) for the ${\cal M}(4,5)$
and ${\cal M}(5,6)$ models:
\ba
 && \chi_{2,k}^{4,5}(q) \= \sum_{\vec{m}=0}^\infty 
 \frac{q^{2 m_1^2+\frac 12 m_2^2 + (4-2k)m_1 + \frac 12 m_2}}
 {(q^2)_{m_1} (q)_{m_2}} \,, \quad k=1,2 \,, \label{qq3} \\
 && \chi_{k,2}^{5,6}(q) \pm \chi_{k,4}^{5,6}(q)
 \= \sum_{\vec{m}=0}^\infty \frac{(\pm 1)^{m_1}\,
 q^{\frac 12 (m_1^2 + m_2^2) +(2-k)m_1 + \frac 12 m_2}}
 {(q^2)_{m_1} (q)_{m_2}} \,, \quad k=1,2\,. \label{qq4}
\ea

\section{Discussion.}

Having a character (linear combination of characters) in the fermionic
form (\ref{F2}), we can rewrite it as a series 
$\chi(q)=\sum_{k=0}^{\infty} \mu_k q^k$,  where the level $k$ admits 
partitioning, $k=\sum_{a=1}^r \sum_{i_a} p_a^{i_a}$, 
into parts of a specific form. 
The interpretation of the $p_a^{i_a}$ as momenta of massless particles
gives rise to the quasi-particle picture, where a character is regarded
as a partition function, $\chi(q)=\sum_k \mu_k e^{-\beta E_k}$. 
Here $q=e^{-2\pi \beta v/L}$, with $v$ being the speed of sound, and 
$L$ -- the size of the system. This quasi-particle representation was
developed originally in \cite{KKMM} (for $b_i=1$) and became a standard
technique by now. It is also applicable to factorized characters 
\cite{BF2,BF1,BF3} (in this case $b_i \neq 1$).

For the fermionic form (\ref{F1}) or (\ref{F2}) of a character the 
quasi-particle representation involves $r$ quasi-particles. They are
naturally interpreted as a conformal limit of particles presented in
a massive theory related to the given conformal model. Moreover, it was 
suggested in \cite{BM} that different non-equivalent fermionic 
representations of the same character correspond to different integrable 
perturbations of the conformal model in question. It would be interesting
to understand if the representations for ${\cal M}(4,5)$, ${\cal M}(5,6)$
and ${\cal M}(6,7)$ discussed above in Section 2 agree with this picture. 
Our results demonstrate that the number of non-equivalent 
fermionic representations for ${\cal M}(s,t)$ increases if $st$ can be
represented as a product of two other co-prime numbers. For instance,
we have encountered above three representations of the type (\ref{F1})
for ${\cal M}(5,6)$ besides the one considered in \cite{KKMM}. Furthermore, 
we can considerably expand the list of non-equivalent fermionic representations, 
if we are looking for representations of the type 
(\ref{F2}), including those that are reducible. For instance, in this way
one obtains representations with one (\ref{q2}) and two quasi-particles
(\ref{qq4}) for ${\cal M}(5,6)$ (another two-particle representation of 
the type (\ref{F2}) follows from the factorized characters \cite{BF2,BF3}).

To summarize, in the present paper we have extended the list of fermionic
representations for some minimal Virasoro models. Physical content of 
these representations, in particular, their connection with massive 
integrable models remains to be investigated. For all considered fermionic 
representations we have established Rogers-Ramanujan type identities and
the corresponding dilogarithm identities. The Rogers-Ramanujan type 
identities possibly can be employed to construct various quasi-particle 
representations for certain physical entities arising in the lattice 
models of statistical mechanics.

\vspace{0.5mm}
After submission of this manuscript I was informed by the referee
that connection of the matrix (\ref{M45a}) to the ${\cal M}(4,5)$ model
was found earlier by M.~Terhoeven (unpublished).

\vspace*{3mm}
\noindent {\bf Acknowledgement:} \
I am grateful to A.~Fring for useful discussions.

\vspace*{-3mm}
\newcommand{\sbibitem}[1]{ \vspace*{-1.5ex} \bibitem{#1}  }

\end{document}